\documentclass[aps,prl,twocolumn,superscriptaddress]{revtex4-1}
\usepackage{amssymb, amsmath}
\usepackage{graphicx,onlyamsmath}

\usepackage{natbib}

\begin{document}

\title{Temperature-driven single-valley Dirac fermions in HgTe quantum wells}

\author{M. Marcinkiewicz}
\affiliation{Laboratoire Charles Coulomb, UMR Centre National de la Recherche Scientifique 5221, University of Montpellier, 34095 Montpellier, France.}

\author{S. Ruffenach}
\affiliation{Laboratoire Charles Coulomb, UMR Centre National de la Recherche Scientifique 5221, University of Montpellier, 34095 Montpellier, France.}

\author{S.~S.~Krishtopenko}
\affiliation{Laboratoire Charles Coulomb, UMR Centre National de la Recherche Scientifique 5221, University of Montpellier, 34095 Montpellier, France.}
\affiliation{Institute for Physics of Microstructures RAS, GSP-105, 603950, Nizhni Novgorod, Russia}

\author{A.~M.~Kadykov}
\affiliation{Laboratoire Charles Coulomb, UMR Centre National de la Recherche Scientifique 5221, University of Montpellier, 34095 Montpellier, France.}
\affiliation{Institute for Physics of Microstructures RAS, GSP-105, 603950, Nizhni Novgorod, Russia}

\author{C. Consejo}
\affiliation{Laboratoire Charles Coulomb, UMR Centre National de la Recherche Scientifique 5221, University of Montpellier, 34095 Montpellier, France.}

\author{D.~B.~But}
\affiliation{Laboratoire Charles Coulomb, UMR Centre National de la Recherche Scientifique 5221, University of Montpellier, 34095 Montpellier, France.}

\author{W. Desrat}
\affiliation{Laboratoire Charles Coulomb, UMR Centre National de la Recherche Scientifique 5221, University of Montpellier, 34095 Montpellier, France.}

\author{W.~Knap}
\affiliation{Laboratoire Charles Coulomb, UMR Centre National de la Recherche Scientifique 5221, University of Montpellier, 34095 Montpellier, France.}
\affiliation{Institute of High Pressure Physics, Polish Academy of Sciences, Soko{\l}owska 29/37 01-142 Warsaw, Poland.}

\author{J.~Torres}
\affiliation{Institut d'Electronique et des Systemes, UMR Centre National de la Recherche Scientifique 5214, University of Montpellier, 34095 Montpellier, France.}

\author{A.~V.~Ikonnikov}
\affiliation{Institute for Physics of Microstructures RAS, GSP-105, 603950, Nizhni Novgorod, Russia}

\author{K.~E.~Spirin}
\affiliation{Institute for Physics of Microstructures RAS, GSP-105, 603950, Nizhni Novgorod, Russia}

\author{S.~V.~Morozov}
\affiliation{Institute for Physics of Microstructures RAS, GSP-105, 603950, Nizhni Novgorod, Russia}
\affiliation{Lobachevsky State University of Nizhni Novgorod, pr. Gagarina 23, 603950 Nizhni Novgorod, Russia.}

\author{V.~I.~Gavrilenko}
\affiliation{Institute for Physics of Microstructures RAS, GSP-105, 603950, Nizhni Novgorod, Russia}
\affiliation{Lobachevsky State University of Nizhni Novgorod, pr. Gagarina 23, 603950 Nizhni Novgorod, Russia.}

\author{N.~N.~Mikhailov}
\affiliation{Institute of Semiconductor Physics, Siberian Branch, Russian Academy of Sciences, pr. Akademika Lavrent'eva 13, Novosibirsk, 630090 Russia}
\affiliation{Novosibirsk State University, Pirogova st. 2, 630090 Novosibirsk, Russia.}

\author{S.~A.~Dvoretskii}
\affiliation{Institute of Semiconductor Physics, Siberian Branch, Russian Academy of Sciences, pr. Akademika Lavrent'eva 13, Novosibirsk, 630090 Russia}
\affiliation{Novosibirsk State University, Pirogova st. 2, 630090 Novosibirsk, Russia.}

\author{F.~Teppe}
\email[]{frederic.teppe@umontpellier.fr}
\affiliation{Laboratoire Charles Coulomb, UMR Centre National de la Recherche Scientifique 5221, University of Montpellier, 34095 Montpellier, France.}
\date{\today}

\begin{abstract}
We report on temperature-dependent magnetospectroscopy of two HgTe/CdHgTe quantum wells below and above the critical well thickness $d_c$. Our results, obtained in magnetic fields up to 16~T and temperature range from 2~K to 150~K, clearly indicate a change of the band-gap energy with temperature. The quantum well wider than $d_c$ evidences a temperature-driven transition from topological insulator to semiconductor phases. At the critical temperature of 90~K, the merging of inter- and intra-band transitions in weak magnetic fields clearly specifies the formation of gapless state, revealing the appearance of single-valley massless Dirac fermions with velocity of $5.6\times10^5$~m$\times$s$^{-1}$. For both quantum wells, the energies extracted from experimental data, are in good agreement with calculations on the basis of the 8-band Kane Hamiltonian with temperature-dependent parameters.
\end{abstract}

\maketitle

Within the last decade, realizations of massless Dirac fermions (DFs) have been extensively studied in condensed matter systems~\cite{Refl1}. This study began with the discovery of graphene hosting two-dimensional (2D) massless DFs coming from two non-equivalent valleys~\cite{Refl2a,Refl2b}. Since then, 2D and 3D massless fermions have also been identified at the surfaces of 3D topological insulators (TIs)~\cite{RefF1} and in Dirac and Weyl semimetals~\cite{RefF2,RefF3,RefF3a,RefF4}. HgTe-based quantum wells (QWs) were the first 2D systems after graphene, in which massless DFs were experimentally demonstrated~\cite{Refl3}. As the QW width $d$ is varied, the first electron-like subband (\emph{E}1) crosses the first hole-like subband (\emph{H}1)~\cite{Refl4a,Refl4b}. When $d$ is smaller than a critical width $d_c$, the \emph{E}1 subband energy is larger than the one of \emph{H}1 subband, and a semiconductor (SC) phase is obtained with a conventional alignment of the electronic states. Above $d_c$, the \emph{E}1 subband drops below the \emph{H}1 subband and the 2D TI phase is formed by this inverted band ordering~\cite{Refl4,Refl5}. Consequently, at the critical thickness $d_c$ the band-gap closes, establishing the topological transition between SC and TI phases, and the QW hosts single-valley 2D massless DFs~\cite{Refl3}.

In addition to the QW thickness, hydrostatic pressure~\cite{X1} and temperature~\cite{Refl6} should also induce the transition from SC to TI phases across the gapless state. By using temperature or pressure as a fine-tuning external parameter one may therefore precisely adjust the QW band-gap to zero and observe the single-valley massless DFs in HgTe QWs. Recently, it has been shown the fingerprints of temperature-induced transition from the TI at 4.2 K to the SC phase at 300~K, measured by magnetotransport up to 30~T~\cite{Refl7}. However, the critical temperature at which the phase transition occurs and the massless DFs are formed, could not be determined by this experimental technique at high temperature. The latter is caused by significant deterioration of resolution between Landau levels (LLs) observed in magnetotransport in 2D systems with increasing of temperature.

One of the specific properties of massless fermions is their behaviour in a perpendicular magnetic field, which transforms a zero-field continuum of electronic states into a set of non-equidistantly spaced LLs with a square-root dependence of their energy on magnetic field~\cite{Refl2a,Refl2b,RefF1,RefF2,RefF3,RefF3a,RefF4}. Recently, the ability to probe temperature-induced 3D massless fermions in HgCdTe crystals by far-infrared (FIR) magneto-absorption spectroscopy was reported, enabling direct and accurate measurements of the fermi velocity~\cite{RefF3a}.

In this letter, we report on the first unambiguous observation of 2D massless DFs induced by temperature in HgTe QWs by FIR magnetospectroscopy. The previous magnetospectroscopy studies of DFs in HgTe QWs have either been performed at low temperatures~\cite{X2,X3,X4,X4a,X4b} or only probed the temperature evolution of LL transitions with monochromatic TeraHertz (THz) light sources~\cite{X5}. Here, by probing the inter- and intra-band LL transitions, we explore the continuous evolution of the band structure with temperature and define a critical temperature $T_c$, corresponding to the arising of 2D massless DFs and to the topological phase transition in HgTe QWs.

\onecolumngrid
\begin{center}
\begin{figure}[h]
\includegraphics [width=0.93\columnwidth, keepaspectratio] {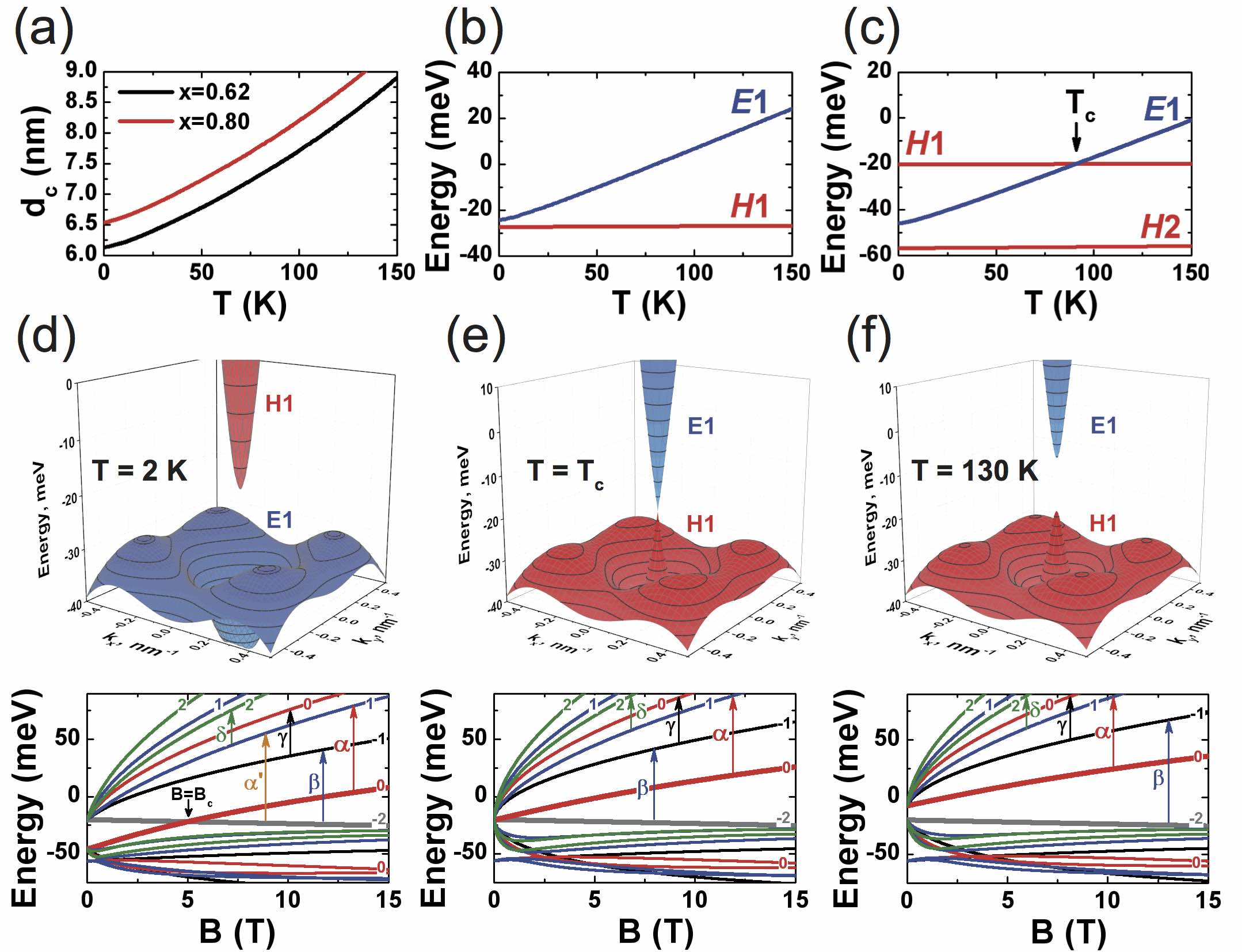} 
\caption{\label{Fig:1} (a) Critical QW width $d_c$, corresponding to the phase transition and arising of massless Dirac fermions, as a function of temperature in (013) HgTe/Cd$_{x}$Hg$_{1-x}$Te QWs for $x=0.62$ and 0.80. (b,c) Temperature dependence of the electron-like \emph{E}1 and the heavy-hole-like \emph{H}1, \emph{H}2 subbands as a function of temperature at zero quasimomentum (b) for sample A and (c) sample B. (d,e,f) Band structure (the top panels) and Landau levels (the bottom panels) in sample B at different temperatures: (d) $T=2$~K, (e) $T=T_c$ and (f) $T=130$~K. The \emph{E}1 subband in the top panels is shown in blue, the red surface corresponds to the \emph{H}1 subband. The $x$ and $y$ axes are oriented along (100) and (0$3\bar{1}$) crystallographic directions, respectively. The numbers over the curves in the bottom panels correspond to the LL indices. A pair of zero-mode LLs with indices $-2$ and $0$ is shown by bold curves. The arrows and Greek letters denote LL transitions, observed in magneto-absorption spectra of sample B.}
\end{figure}
\end{center}
\twocolumngrid

The two QW samples studied in this work were grown by molecular beam epitaxy on [013]-oriented semi-insulating GaAs substrate with relaxed CdTe buffer~\cite{Ref1}, with nominal well widths $d$ of 6~nm (sample A) and 8~nm (sample B). The HgTe QW is embedded in Cd$_x$Hg$_{1-x}$Te barriers with a nominal thickness of about 40~nm, $x= 0.62$ for sample A and 0.80 for sample B. A CdTe cap layer was deposited on top of the structures. The QW in sample B is remotely doped on each side by a 15-nm-thick In-doped region with the doping concentration of $6.5\times10^{16}$ cm$^{-3}$ resulting in the electron concentration in the well of $3\times10^{11}$ cm$^{-2}$ at 2~K. The samples A is nominally undoped with the concentration of 2D holes of $3\times10^{10}$ cm$^{-2}$ at the low temperature.

According to the temperature-dependent band structure calculations on the basis of the 8-band Kane Hamiltonian~\cite{X1}, sample A is expected to be almost gapless but with direct band ordering at 2~K and its gap increases with temperature. It is clearly seen in Fig.~\ref{Fig:1} that the band ordering in sample B changes with temperature. The critical temperature $T_c$, corresponding to the phase transition, is estimated to be 90~K.

The origin of temperature-driven band ordering in HgTe/CdTe QWs is caused by strong temperature dependence of the energy gap at the $\Gamma$ point between the $\Gamma_6$ and $\Gamma_8$ bands in HgCdTe crystals~\cite{RefF3a}. Since the band gap in HgTe/CdTe QW depends on quantum confinement and, consequently, on the energy gap difference of the well and barrier materials, variation of both temperature and QW width influences the band ordering in the QW. We note that for quantitative description of the temperature effect on the band ordering, additionally to the gap, one should also take into account the temperature dependence of the valence band offset, the lattice constants and the elastic constants in the bulk materials~\cite{X1}.

Fig.~\ref{Fig:1} shows that at 2~K sample B is TI with indirect-gap of about 10~meV, arising due to the presence of four side maxima in the valence band. At the transition point, the linear band dispersion in the vicinity of the $\Gamma$ point of the Brillouin zone clearly features the presence of massless DFs. At $T>T_c$, sample B is a normal direct-gap semiconductor. The bottom panels in Fig.~\ref{Fig:1} present the LL fan chart at 2~K, 90~K and 120~K. To calculate LLs, we imply the axial approximation~\cite{X1} by keeping the in-plane rotation symmetry and omitting the warping terms and also the terms resulting from bulk inversion asymmetry (BIA) of the unit cell in bulk zinc-blende crystals. In this case the electron wave-function for a given LL index $n>0$  has generally eight components, describing the contribution of the $\Gamma_6$, $\Gamma_7$ and $\Gamma_8$ bands into the LL. We note that specific LL with $n=-2$ contains only contribution of the heavy-hole band with momentum projection $-3/2$~\cite{X2,X3,X1a}. Details of the LL notation can be found in~\cite{X1}.

\begin{figure}
\includegraphics [width=1.0\columnwidth, keepaspectratio] {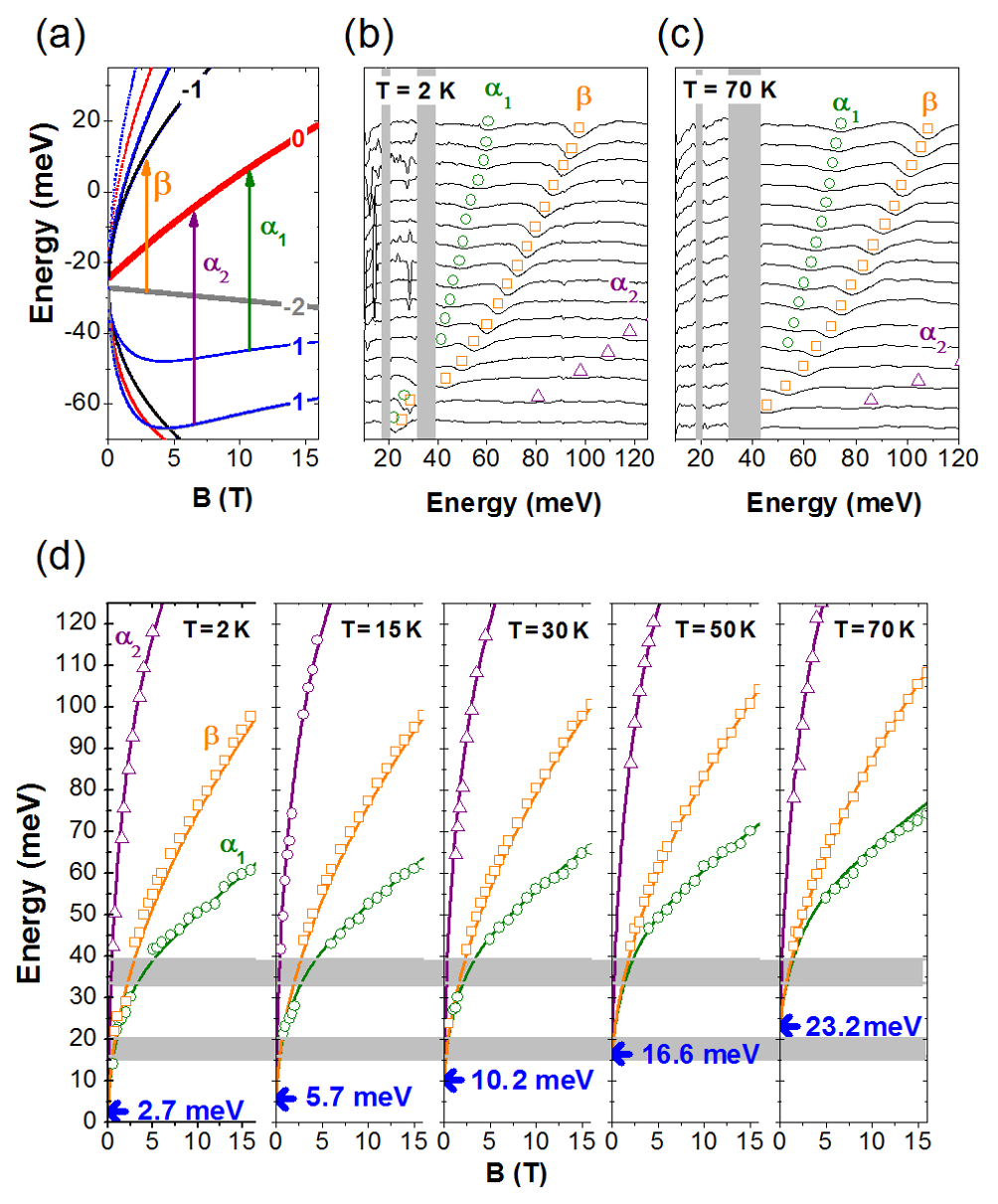} 
\caption{\label{Fig:3} (a) LLs as a function of magnetic field in sample A at 2~K. The numbers conform to the LL indices. The zero-mode LLs are shown by bold curves. The arrows and Greek letters denote LL transitions, observed in magneto-absorption spectra. (b,c) Transmission spectra of sample A at 2~K and 70~K from 1~T (the first plot at the bottom) to 16~T (the last plot on the top) with the step for magnetic field of 1.0~T. (d) Energy of $\alpha_1$ (green curve), $\alpha_2$ (purple curve) and $\beta$ (orange curve) transitions as a function of magnetic field. The experimental data are represented by symbols: circles, triangles and squares for the $\alpha_1$, $\alpha_2$ and $\beta$ transitions, respectively. The value of the band-gap is shown by blue arrows. Shaded areas indicate the reststrahlen bands, although in order to simplify the figure, their evolution with temperature has not been represented here. For simplicity, evolution of the reststrahlen bands with temperature is not shown in the panel.}
\end{figure}

The inherent property of each phase is characterized by the behavior of a particular pair of LLs, so-called zero-mode LLs, under applied magnetic field $B$~\cite{Refl3,Refl4,Refl5}. The origin of this peculiar pair of LLs becomes apparent when using a modified $4\times4$ Dirac-type Hamiltonian~\cite{Refl4} for the approximate description of electronic states at small values of quasimomentum $\textbf{k}$. For the inverted band structure, below a critical field value $B_c$, the lowest zero-mode LL has electron-like character and arises from the valence band, while the highest zero-mode LL has a heavy-hole-like character and splits from the conduction band. With increasing magnetic field, the zero-mode LLs cross each other at $B=B_c$. For the direct band ordering, the zero-mode LLs are not crossed as the electron- and heavy-hole-like level arises at $B=0$ in conduction and valence band, respectively. Such particular pair of the zero-mode LLs is defined by the LLs with $n=-2$ and $n=0$ and can be easily recognized in Fig.~\ref{Fig:1}. In contrast, sample A remains to be in the SC phase with temperature increasing, and, therefore, their zero-mode LLs do not cross with magnetic field (see Fig.~\ref{Fig:3}a).

To probe temperature evolution of the zero-mode LLs, we have performed FIR magneto-transmission measurements in the Faraday configuration by using a Fourier transform spectrometer coupled to a 16~T superconducting coil. Specific design of the experimental setup~\cite{RefF3a} allows continous tuning of temperature in the range from 2~K up to 150~K. The magneto-absorption spectra were measured with a spectral resolution of 0.75~meV. All spectra were normalized by the sample transmission at zero magnetic field.

The magnetotransmission spectra of sample A at 2~K and 70~K are presented in panel (b) and (c) of Fig.~\ref{Fig:3}. The spectra for other temperatures are given in~\cite{SM}. Samples are completely opaque in the range of the Cd$_x$Hg$_{1-x}$Te and the GaAs reststrahlen bands, indicated by shaded areas. In Faraday configuration, optically active inter-LLs transitions follow the conventional selection rules $\Delta n=\pm1$ (for unpolarized radiation) imposed by the electric dipole approximation. Due to the small hole concentration in sample~A, only few LLs in valence band are occupied. For instance, $B\simeq1.2$~T corresponds to LL filling factor $\nu=1$ for the concentration at 2~K. Therefore, three high-intense lines in the spectra for all temperatures can be identified as inter-band LL transitions, involving the zero-mode LLs. Those transitions are marked in Fig.~\ref{Fig:3}a with small Greek letters, in accordance with the previously used notations~\cite{X2,X3,X4,X4a}.

Comparison between experimental and theoretical values of the transition energies is presented in Fig.~\ref{Fig:3}d. We note that extrapolation of the energy behaviour in magnetic field of the inter-band LL transitions into $B=0$ allows evaluating the band-gap at zero quasimomentum. A very good agreement of experimental data with theoretical calculations clearly demonstrates the band-gap opening in sample A, induced by temperature. The latter is indicated in Fig.~\ref{Fig:3}d by blue arrows.

\begin{figure}
\includegraphics [width=1.0\columnwidth, keepaspectratio] {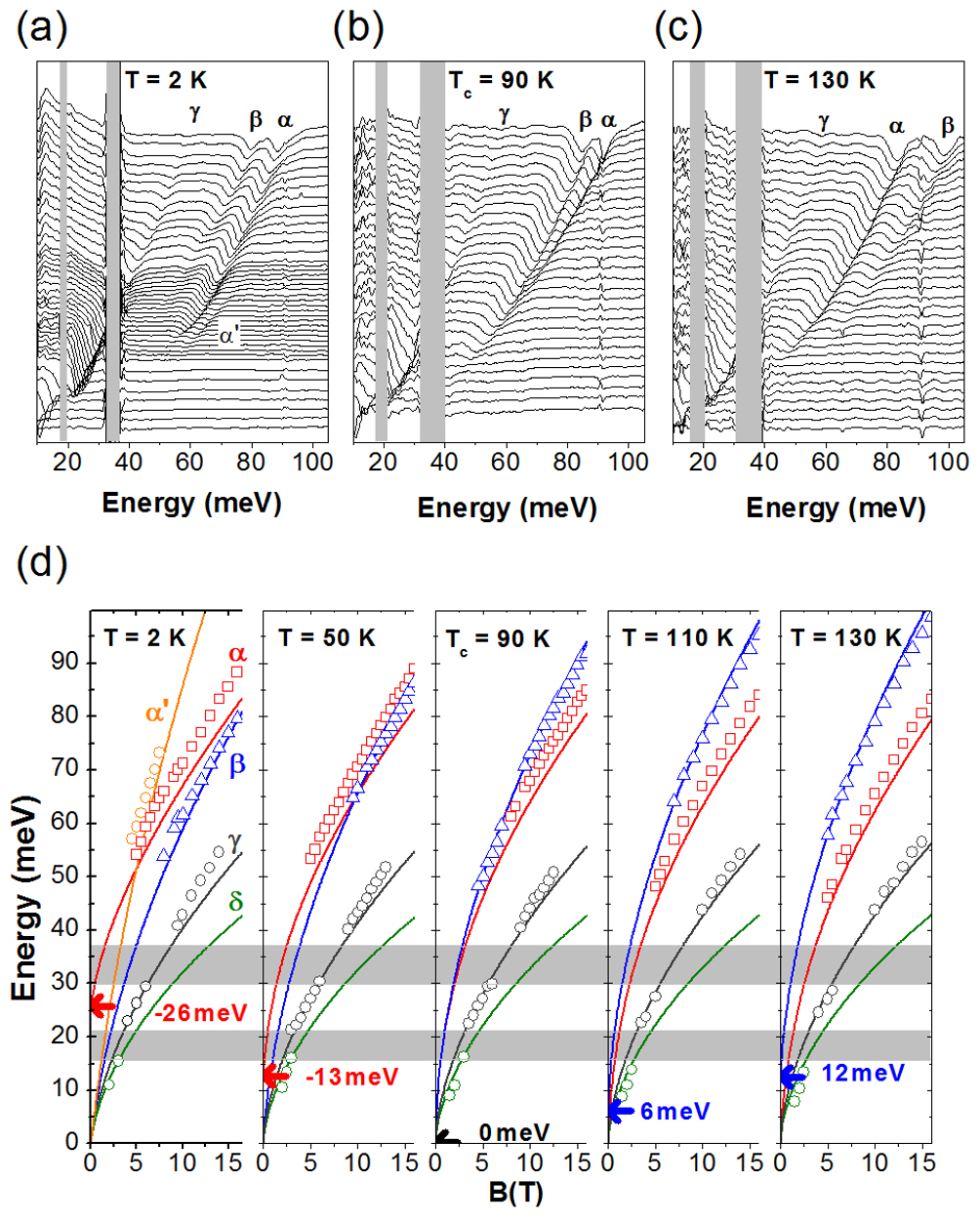} 
\caption{\label{Fig:4} (a,b,c) Transmission spectra of sample B at 2~K, 90~K and 130~K from 0.5~T (the first plot at the bottom) to 16~T (the last plot on the top). The step for magnetic field is 0.5~T, except the panel for 2~K, in which a 0.25~T step was used between 4~T and 9~T allowing to accurately follow the evolution of $\alpha'$ transition. The Greek letters denote LL transitions, shown in Fig.~\ref{Fig:1}. (d) Fan chart of inter-LL transitions in sample B. The $\alpha'$, $\alpha$, $\beta$, $\gamma$ and $\delta$ calculated transitions are shown in solid lines in orange, red, dark grey and green respectively. Experimental data are represented by symbols in the same colors as the theoretical curves. The gap at $k=0$ is shown by blue and red arrows respectively for negative and positive values, and black for the gapless state. Shaded areas indicate the reststrahlen bands, although in order to simplify the figure, their evolution with temperature has not been represented here. For simplicity, evolution of the reststrahlen bands with temperature is not shown.}
\end{figure}

Figure~\ref{Fig:4} presents magnetotransmission spectra of sample B at 2~K, 90~K and 130~K (the top panels). The spectra for other temperatures are given in~\cite{SM}. The three most intense lines observed at all temperatures are identified as the LL transitions from the zero-mode LLs ($\alpha$ and $\beta$ transitions) and cyclotron resonance (CR) absorption due to 2D electrons ($\gamma$ and $\delta$ transitions). By using the electron concentration at 2~K, one concludes that CR in the fields below 3.4 T, corresponding to $\nu\approx3$, can be related with both $\gamma$ and $\delta$ transitions, while above this field only $\gamma$ transition is possible. It is seen that the $\gamma$ line disappears from the spectrum at $B\approx12$~T. Assuming that $B = 12$~T corresponds to $\nu=1$, we find an excellent agreement upon the electron concentration, previously derived from magnetotransport measurements.

At 2~K, an additional $\alpha'$ transition, shown in Fig.~\ref{Fig:1} appears in the spectra in a relatively narrow range of magnetic fields $B=5-–7.5$~T, in which the zero-mode LLs are crossed. Such transition does not satisfy the selection rules $\Delta n=\pm1$ and is forbidden in the electric dipole approximation. The arising of this transition, also previously reported in \cite{X2,X3,X4}, is related with the coupling between the $n=0$ and $n=2$ LLs, resulting from BIA, initially neglected in our calculations. Taking into account BIA leads to the anticrossing in the vicinity of $B=B_c$ and to the mixing of the states at the zero-mode LLs. The latter makes the $\alpha'$ transition to become active in the electric dipole approximation.
Observation of the $\alpha'$ line, which is followed up to 30~K in our data, is direct evidence of the inverted band structure of sample B at low temperatures.

As it has been mentioned above, a distinctive characteristic of massless particles is the square-root dependence of energies of LL transitions on magnetic field~\cite{Refl2a,Refl2b,RefF1,RefF2,RefF3,RefF3a,RefF4}. However, linear subband dispersion in HgTe QWs exists only in the vicinity of the $\Gamma$ point. At large values of quasimomentum $k$, the terms, proportional to $k^2$, in the Hamiltonian become relevant~\cite{Refl4,Refl5}. The latter gives rise to the square-root behaviour in weak magnetic fields only, while at high magnetic fields, the linear dependence should be seen. This can be shown by straightforward calculations on the basis of the 8-band Kane Hamiltonian~\cite{SM}.

A more representative characteristic of gapless state in HgTe QWs is the behaviour of the transitions from the zero-mode LLs. If the band structure is inverted, the energy of $\alpha$ transition leans towards the gap energy at $k=0$ when $B$ tends to zero, while the energy of $\beta$ transition vanishes. For direct band ordering, the transition behavior in weak magnetic fields is reversed: the gap at $k=0$ corresponds to the cut-off energy for the $\beta$ transition, however, the energy of the $\alpha$ transition tends to zero. In the gapless state with 2D massless DFs, the energy of both transitions should have the same dependence on $B$ in weak magnetic fields, and the corresponding absorption lines merge if magnetic field goes to zero. The latter is seen in the transmission spectra at 90~K that indicates a vanishing of the gap at $k=0$ and, hence, arising of 2D massless DFs.

We note that merging of $\alpha$ and $\beta$ lines with decreasing of magnetic field in the  gapless state can also be derived analytically from a modified $4\times4$ Dirac-type 2D Hamiltonian~\cite{Refl4}, also used for the description of previous magnetotransport results on 2D massless DFs~\cite{Refl3}. Unfortunately, the large number of variable parameters of this 2D model does not allow to use it for efficient fitting of the experimental data. As a result, only the observation of the $\alpha$ and $\beta$ LL transitions behavior in transmission spectra makes possible to demonstrate the band-gap vanishing and to determine the critical temperature $T_c$, while the band velocity $v_F$ of massless DFs can not be directly extracted from the data. However, a good agreement between theoretical calculations and experimental data for sample B evidences that actual band velocity of massless DFs at 90~K should be very close to the theoretical value $v_F=5.6\times10^5$ m$\times$s$^{-1}$~\cite{SM}.

In conclusion, we have demonstrated the ability to observe the changing of the band-gap in HgTe QWs by temperature-dependent FIR magnetotransmission spectroscopy. In the case of the inverted band structure, we have determined a critical temperature $T_c=90$~K, corresponding to the band-gap vanishing and, hence, arising of single-valley 2D massless DFs. A good agreement between experimental results and theoretical calculations on the basis of the 8-band Kane Hamiltonian with temperature-dependent parameters allows us to evaluate the band velocity of 2D massless DFs.
~~~\\
\begin{acknowledgments}
This work was supported by the CNRS through LIA TeraMIR project, by the Languedoc-Roussillon region via the "Gepeto Terahertz platform", by Eranet-Rus-Plus European program "Terasens", by Russian Academy of Sciences, by Russian Foundation for Basic Research (Grant Nos. 15-52-16017 and 16-02-00672) and by Russian Ministry of Education and Science (MK-1136.2017.2). Theoretical calculations and characterization of the samples were performed in the framework of project 16-12-10317 provided by Russian Science Foundation.
\end{acknowledgments}


%

\newpage
\clearpage
\setcounter{equation}{0}
\setcounter{figure}{0}
\setcounter{table}{0}


\onecolumngrid
\section*{Supplemental Materials}
\maketitle
\onecolumngrid
\subsection{Dirac-type 2D Hamiltonian}
To describe qualitatively the band inversion in HgTe/Cd(Hg)Te QWs, one can also use the effective Dirac-type 2D Hamiltonian~\cite{w1}, proposed for the electronic states in \emph{E}1 and \emph{H}1 subbands in the vicinity of the $\Gamma$ point of the Brillouin zone. Using the states $|E1,+\rangle$, $|H1,+\rangle$, $|E1,-\rangle$, $|E1,-\rangle$ as a basis, the Hamiltonian for the E1 and H1 subbands is written as follows

\begin{equation}\label{eq:SM1}
  \hat{H}_{eff}(k_x,k_y) = \begin{pmatrix}
    \hat{H}_{D}(\textbf{k}) & 0 \\
    0 & \hat{H}_{D}^{*}(-\textbf{k})
  \end{pmatrix},
\end{equation}
where
\begin{equation}\label{eq:SM2}
\hat{H}_{D}(\textbf{k})=\epsilon(\textbf{k})+\sum_{i=1}^3 d_i(\textbf{k})\sigma_i,
\end{equation},
\begin{equation*}
d_1+id=A(k_x+ik_y)=Ak_{+},~~~~d_3=M-B(k_x^2+k_y^2),~~~~\epsilon=C-D(k_x^2+k_y^2).
\end{equation*}
Here, $k_x$ and $k_y$ are momentum components in the QW plane, and $A$, $B$, $C$ and $D$ are specific QW constants, being defined by QW geometry, material parameters and temperature. The two components of the Pauli matrices $\sigma_i$ denote the \emph{E}1 and \emph{H}1 subbands, whereas the two diagonal blocks $\hat{H}_D(\textbf{k})$ and $\hat{H}_{D}^{*}(-\textbf{k})$ represent spin-up and spin-down states, linked together by time-reversal symmetry. Here, as in the main text, we have neglected the terms, arising due to the bulk inversion asymmetry (BIA) in the unit cell of zinc-blend materials~\cite{w6}, and the terms resulting from the inversion asymmetry of HgTe/CdTe interface~\cite{w3}.

The most important quantity in $\hat{H}_{eff}(k_x,k_y)$ is the mass parameter $M$, which describes the ordering of \emph{E}1 and \emph{H}1 subbands. At the critical temperature $T=T_c$, the mass parameter is equal to zero. If we then only keep the terms up to linear order in $\textbf{k}$ for each spin, $\hat{H}_{D}^{*}(-\textbf{k})$ and $\hat{H}_{D}(\textbf{k})$ correspond to Hamiltonians, describing massless Dirac fermions. As it has no valley degeneracy, HgTe/Cd(Hg)Te QWs with $M=0$ offer realization of single-valley massless Dirac fermions~\cite{w2}. In this case, parameter $A$, describing the non-diagonal terms in $\hat{H}_D(\textbf{k})$ and $\hat{H}_{D}^{*}(-\textbf{k})$, defines the velocity $v_F$ of massless particles.

The negative values of $M$ correspond to the inverted band structure, while $M>0$ describes the direct band ordering. By using the 8-band Kane Hamiltonian, accounting interaction between the $\Gamma_6$, $\Gamma_8$ and $\Gamma_7$ bands in zinc-blend materials, with temperature-dependent parameters~\cite{w4} and by applying the procedure, described in~\cite{w5}, we have calculated the values of $A$, $B$, $C$, $D$ and $M$, as well as $v_F$ at different temperatures (see Table~\ref{tab:1}).

We note that description of electronic states in \emph{E}1 and \emph{H}1 subbands by means of  $\hat{H}_{eff}(k_x,k_y)$ is valid only in the vicinity of the $\Gamma$ point, while the states at large values of $\textbf{k}$ require more realistic approximation on the basis of the 8-band Kane Hamiltonian. Comparison between calculations of subband dispersion in sample B (the 8~nm wide HgTe/Cd$_{0.8}$Hg$_{0.2}$Te QW), performed within the two approaches at different temperatures, is presented in Fig.~\ref{Fig:3SM}.

\subsection{Landau level transitions of massless Dirac fermions in HgTe QWs}

\begin{figure}
\includegraphics [width=0.9\columnwidth, keepaspectratio] {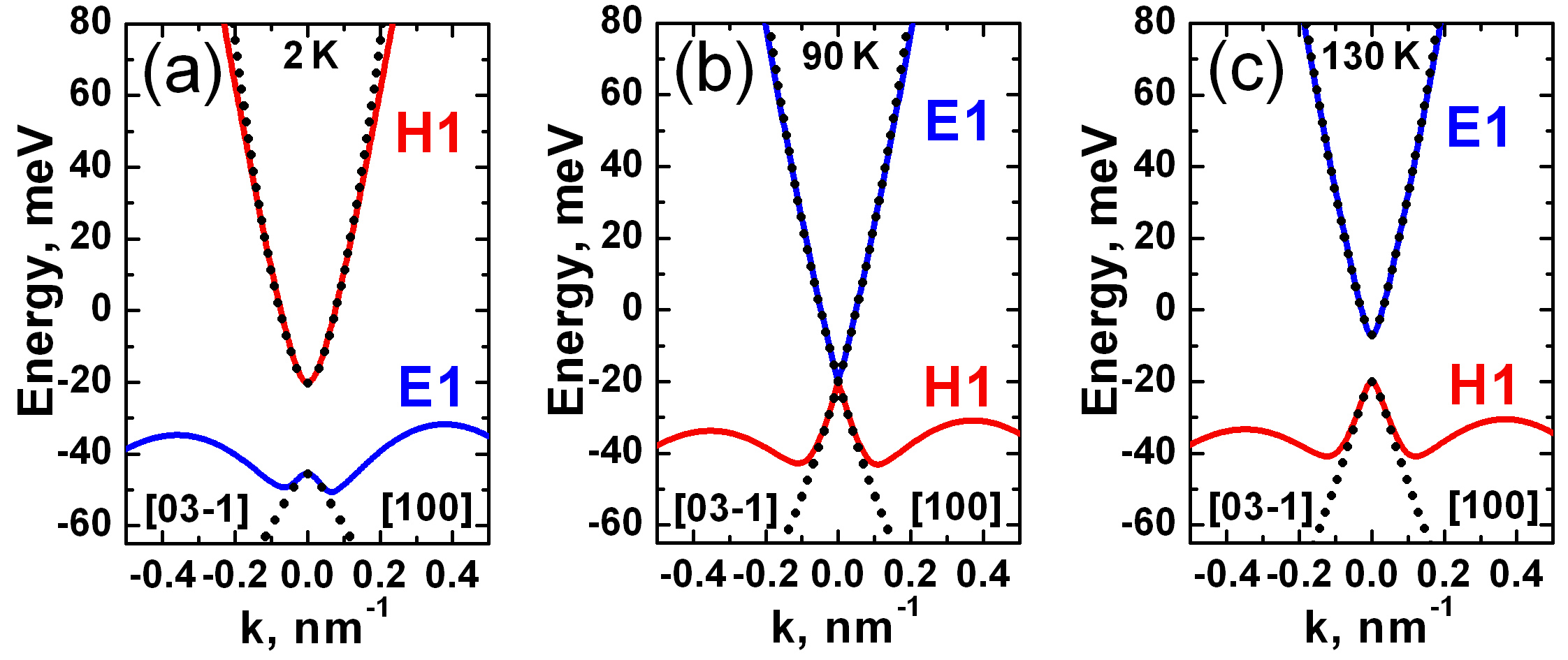} 
\caption{\label{Fig:3SM} Comparison between band structure calculations for sample B, performed by using the 8-band Kane model (red and blue solid curves) and the Dirac-type 2D Hamiltonian (black dotted curves) for different temperatures: (a) $T=2$~K, (b) $T=T_c=90$~K and (c) $T=130$~K.}
\end{figure}

A distinctive characteristic of massless particles with linear band dispersion is a square-root dependence of energies of LL transitions on magnetic field~\cite{w9,w10,w7,w8}. However, as it is seen in Fig.~\ref{Fig:3SM}, linear subband dispersion in HgTe QWs at $T_c$ exists only in the vicinity of the $\Gamma$ point. At large values of quasimomentum $k$, the terms proportional to higher order of $k$ in the Hamiltonian become relevant. For an example, in the  Dirac-type 2D Hamiltonian $\hat{H}_{eff}(k_x,k_y)$~(\ref{eq:SM1}), they are presented by  $B(k_x^2+k_y^2)$ and $D(k_x^2+k_y^2)$ terms.

\begin{table}
\caption{\label{tab:1} Parameters involved in the Dirac-type 2D Hamiltonian $\hat{H}_{eff}(k_x,k_y)$ at different temperatures.}
\begin{ruledtabular}
\begin{tabular}{ccccccc}
Temperature~(K) & $C$~(meV) & $M$~(meV) & $B$~(meV$\cdot$nm$^2$) & $D$~(meV$\cdot$nm$^2$) & $A$~(meV$\cdot$nm) & $v_F$~(m$\cdot$s$^{-1}$) \\
\hline
2 & -32.8 & -12.7 & -993.5 & -810.4 & 353.3 & $5.37\cdot 10^{5}$ \\
90 & -19.9 & 0 & -805.3 & -622.2 & 368.6 & $5.60\cdot 10^{5}$ \\
130 & -13.5 & 6.4 & -733.5 & -550.4 & 378.8 & $5.75\cdot 10^{5}$ \\
\end{tabular}
\end{ruledtabular}
\end{table}

\begin{figure}
\includegraphics [width=0.68\columnwidth, keepaspectratio] {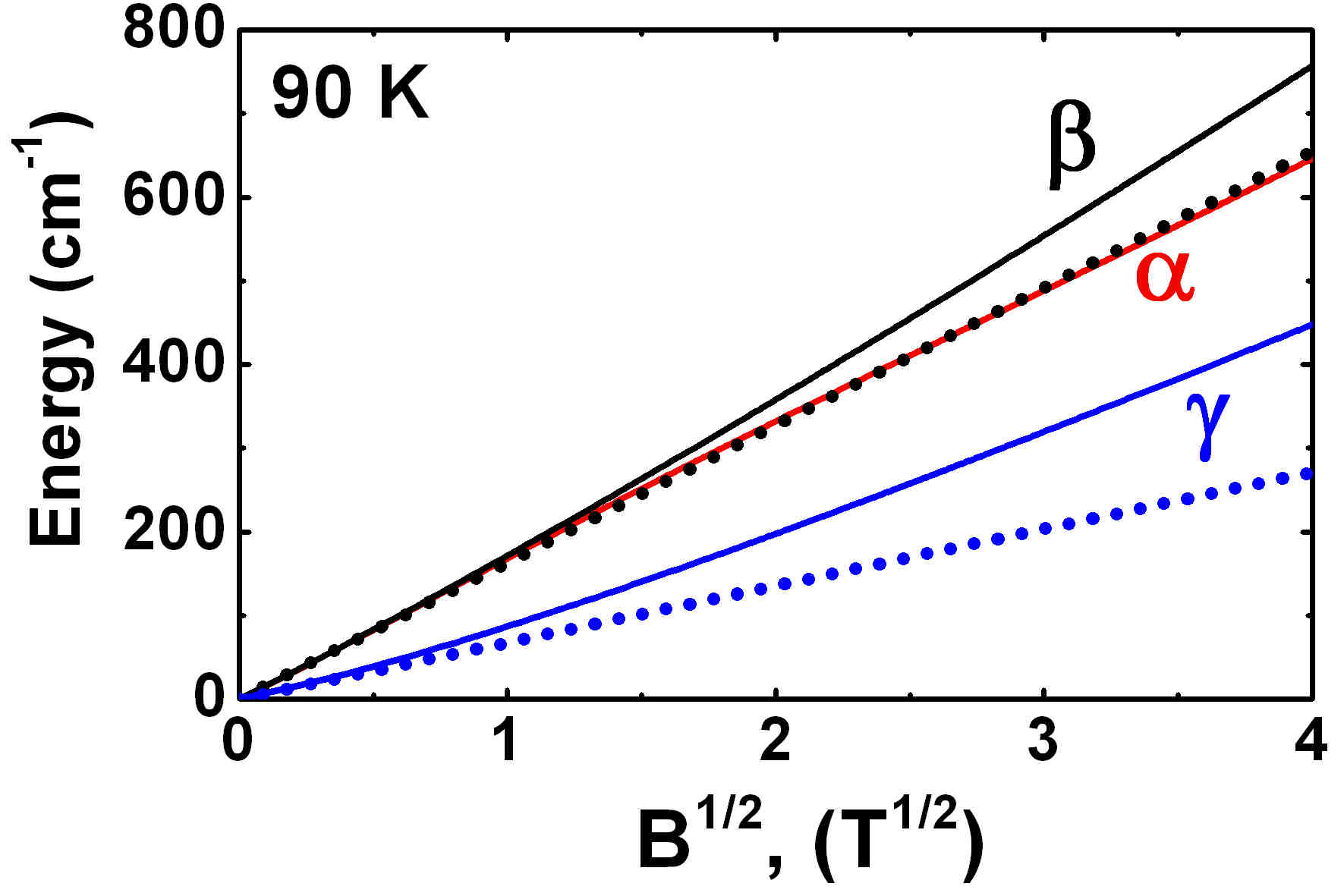} 
\caption{\label{Fig:4SM} Energy evolution of $\alpha$, $\beta$, and $\gamma$ transitions  as a function of $\sqrt{B}$ at $T$ = 90 K calculated within the pure Dirac model (dotted curves) and by using the 8-band Kane Hamiltonian (solid curves).}
\end{figure}

The presence of these non-linear terms gives rise to the square-root behaviour of LL transitions in weak magnetic fields only, while at high magnetic fields, more complex dependence should be seen. To demonstrate it, we provide magnetic field dependence for energies of $\alpha$, $\beta$ and $\gamma$ transitions in sample B for 90~K, at which the massless Dirac fermions arise (see Fig.~\ref{Fig:4SM}). The solid curves are calculations, performed by using the 8-band Kane Hamiltonian. For clarity, we also provide the energies of $\alpha$, $\beta$ and $\gamma$ transitions, described within the pure Dirac model by neglecting of $B(k_x^2+k_y^2)$ and $D(k_x^2+k_y^2)$ terms in the Dirac-type 2D Hamiltonian~(\ref{eq:SM1}). In this case, one recovers the familiar LL fan chart for massless Dirac fermions~\cite{w10}, $E_n=$sgn$(n)\sqrt{2n}A/l_B$ with the band velocity $\hbar v_F=A$ (see Table~\ref{tab:1}). Here, $l_{B}^{2}=\frac{\hbar c}{e\mathbb{B}}$ is the magnetic length and $\mathbb{B}$ is the strength of magnetic field. We note that under such approximation, the $\alpha$ and $\beta$ transitions coincide due to electron-hole symmetry of the model.

Fig.~\ref{Fig:4SM} shows that $\beta$ and $\gamma$ LLs transitions do not follow a square-root magnetic field dependence over the whole range of studied magnetic fields, up to 16 T, even in the presence of the Dirac cone in the vicinity of the $\Gamma$ point. On the contrary, $\alpha$ transition behaves almost like a square root. However, this behavior is not related to the conical band dispersion, but results from the mutual compensation of the high order terms in $k$ in the 8-band Kane model.

A representative characteristic of the gapless state in HgTe QWs is the merging of the $\alpha$ and $\beta$ transitions at low magnetic fields exhibiting the same square-root dependence on $\mathbb{B}$. Consequently, the corresponding absorption lines in the transmission spectra are merging with a square-root behavior when $\mathbb{B}$ tends to zero.

\subsection{THz magneto absorption measurements}
\begin{figure}
\includegraphics [width=0.8\columnwidth, keepaspectratio] {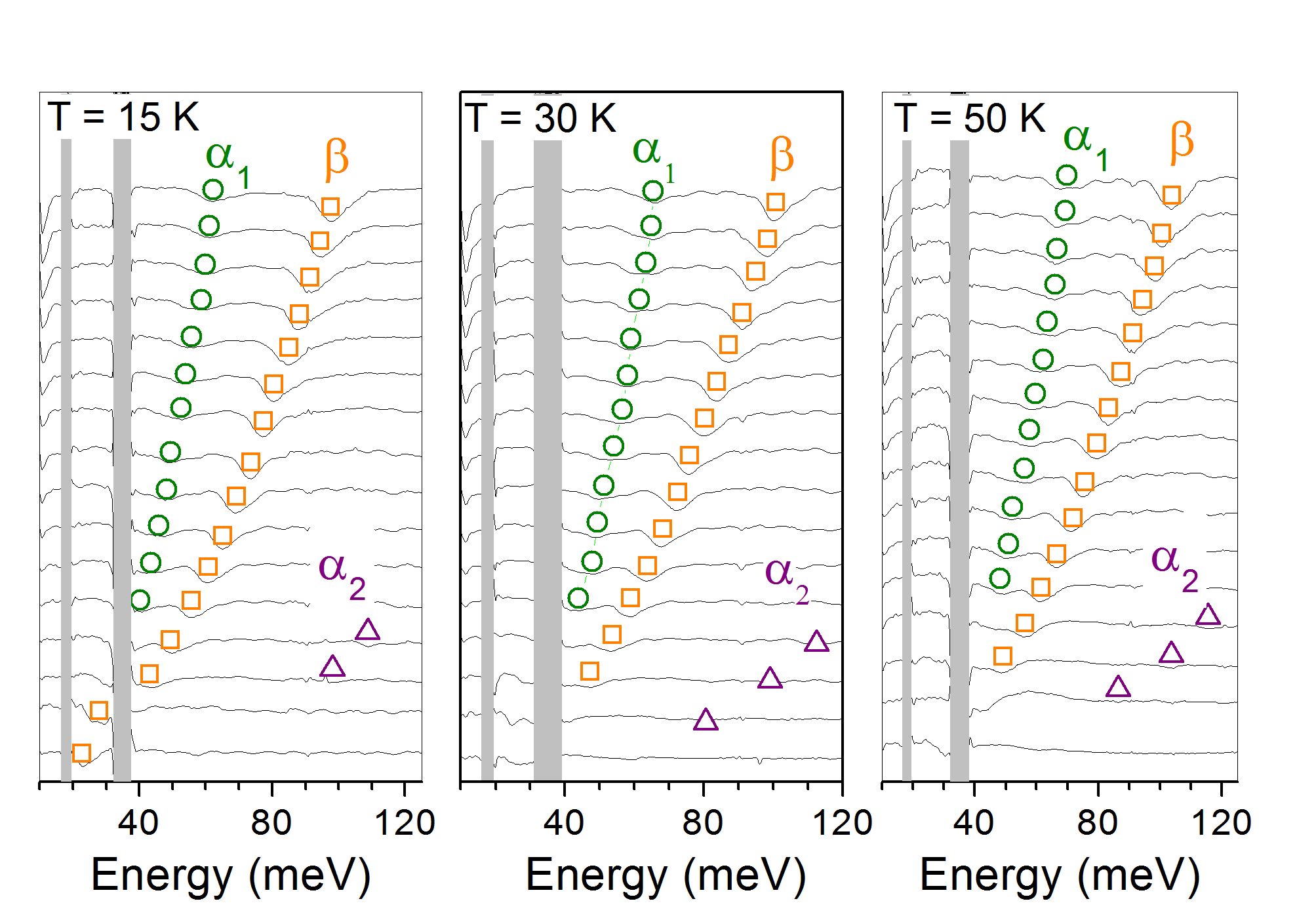} 
\caption{\label{Fig:1SM} Transmission spectra of sample A at 15~K, 30~K and 50~K from 1~T (the first plot at the bottom) to 16~T (the last plot on the top) with the step for magnetic field of 1.0~T.}
\end{figure}

\begin{figure}
\includegraphics [width=0.8\columnwidth, keepaspectratio] {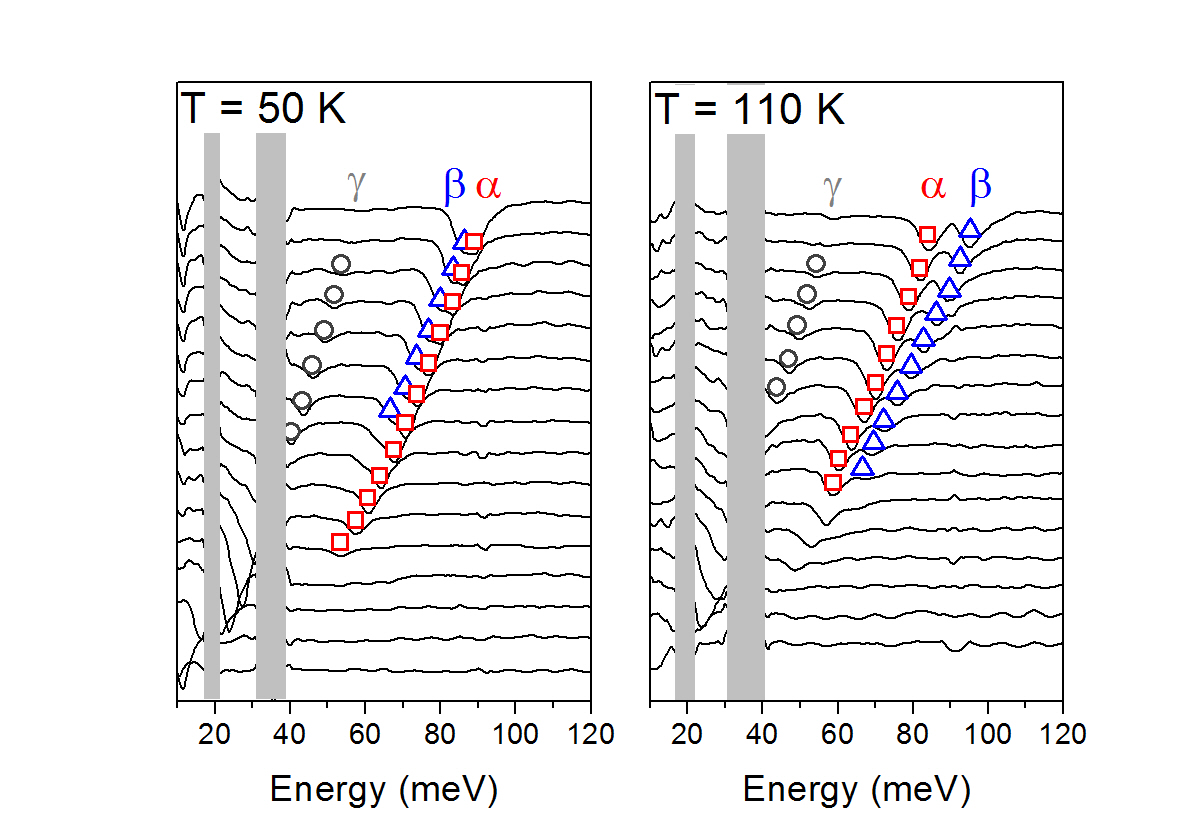} 
\caption{\label{Fig:2SM} Transmission spectra of sample B at 50~K and 110~K from 1.0~T (the first plot at the bottom) to 16~T (the last plot on the top) with the step for magnetic field of 1.0~T.}
\end{figure}

In the main text, we have provided a comparison between experimental results and theoretical calculations for sample A at 2~K, 15~K, 30~K, 50~K, 70~K and sample B at 2~K, 50~K, 90~K, 110~K, 130~K. Here, we provide transmission spectra of sample A (see Fig.~\ref{Fig:1SM}) and sample B (see Fig.~\ref{Fig:2SM}) at the temperatures, which are not presented in the main text.


%

\end{document}